\documentclass{aa}

\usepackage{mathptmx}
\usepackage{graphicx}   
\usepackage{txfonts}
\usepackage{amsmath}    
\usepackage{amssymb}    
\usepackage{longtable}
\usepackage{color}
\usepackage{times}
\usepackage{multirow}

\newcommand{\NObs}{5412}
\newcommand{\NBeam}{6034} 
\newcommand{\us}{$\mu$s}

\newcommand{\psra}{PSR J2205$+$6012}
\newcommand{\psrb}{PSR J2048$+$4951}

\begin{document}

\title{The SPAN512 mid-latitude pulsar survey}

    \subtitle{at the Nançay Radio Telescope}

   \author{G.~Desvignes\inst{1}\fnmsep\inst{2}\fnmsep\thanks{Email: gdesvignes@mpifr-bonn.mpg.de}
          \and
          I.~Cognard\inst{3}\fnmsep\inst{4}
          \and
          D.~A.~Smith\inst{5}\fnmsep\inst{6}
          \and 
          D.~Champion\inst{1}
          \and 
          L.~Guillemot\inst{3}\fnmsep\inst{4}
          \and
          M.~Kramer\inst{1}
          \and
          P. Lespagnol\inst{4}
          \and
          F.~Octau\inst{3}
          \and
          G.~Theureau\inst{3}\fnmsep\inst{4}\fnmsep\inst{7}
          }

\institute{Max-Planck-Institut f\"{u}r Radioastronomie, Auf dem H\"{u}gel 69, D-53121 Bonn, Germany
         \and
             LESIA, Observatoire de Paris, Universit\'e PSL, CNRS, Sorbonne Universit\'e, University Paris Diderot, Sorbonne Paris Cit\'e, 5 place Jules Janssen, 92195 Meudon, France
        \and Laboratoire de Physique et Chimie de l’Environnement et de l’Espace, Universit\'e d’Orl\'eans/CNRS, F-45071 Orl\'eans Cedex 02, France
        \and 
            Observatoire Radioastronomique de Nan{\c c}ay, Observatoire de Paris,
Universit{\'e} PSL, Universit{\'e} d’Orl{\'e}ans, CNRS, 18330 Nan{\c c}ay,
France
        \and
            Centre d'\'Etudes Nucl\'eaires de Bordeaux Gradignan, IN2P3/CNRS, Universit\'e Bordeaux, BP120, 33175 Gradignan, France
        \and
        Laboratoire d'Astrophysique de Bordeaux, Universit\'e Bordeaux, B18N, all\'ee Geoffroy Saint-Hilaire, 33615 Pessac, France
        \and
            LUTH, Observatoire de Paris, PSL Research University, Meudon, France
    }

\date{Received date / Accepted date }

  \abstract
      { The large number of ongoing
  surveys for pulsars and transients at various radio observatories is motivated by
  the science obtained from these sources. Timing and
  polarisation analysis of relativistic binaries can place strong
  constraints on theories of gravity. The observation of a growing number of
  millisecond pulsars (MSPs) spread over the celestial sphere may allow the
  detection of a stochastic gravitational wave background arising from
  supermassive black hole binaries. A more complete sample of young pulsars improves our knowledge of neutron star birth and evolution. Transients such as fast radio bursts can serve to probe the intergalactic medium.}
      { The SPAN512 pulsar survey covers intermediate Galactic latitudes using the L-band receiver of the Nan\c
        cay Radio Telescope (NRT). The survey covers 224 sq. deg. of the sky for a total exposure time of 2200 h. Population syntheses predict the discovery of 3 to
        19 new normal pulsars and a few MSPs.}
     { We present detailed modelling of the NRT beam with its L-band receiver and its sensitivity which we used to precisely assess the expected survey yield. We used the flexible  Pulsar Arecibo L-band Feed Array data processing pipeline to search the 47 TB of SPAN512 data for pulsars and transients.}
{ The SPAN512 survey discovered two new MSPs and one new middle-aged
  pulsar.  
  We focus on the analysis of the 2.4-ms spin period pulsar J2205$+$6012 for which we also report the detection of gamma-ray pulsations. Its narrow pulse width ($35\mu$s at an observing frequency of 2.55~GHz) allows for sub-microsecond timing precision
  over 8 years, with exciting prospects for pulsar
  timing array programs.}
   {}

   \keywords{ (Stars:) pulsars: individual  PSR J2205+6012 --  Stars: neutron -- gamma rays: star -- telescopes  }

   \maketitle

\section{Introduction}
Pulsars are highly magnetised neutron stars with spin periods ranging
from a few milliseconds to a few tens of seconds that emit radio waves
seen as a succession of radio pulses for an observatory located on the
Earth.  Since the fortuitous discovery of the first pulsar
\citep{hbp+68}, numerous radio surveys carried out in the last four
decades have unveiled over 3200 pulsars \citep[as reported by the ATNF
  Pulsar Catalogue
  v1.66\footnote{https://www.atnf.csiro.au/research/pulsar/psrcat/}
  early 2022;][]{mhth05}, with about 35\% discovered by the successful
Parkes Multibeam Pulsar Survey \citep[PMPS;][]{mlc+01}.  Of these,
around $400$ in the Galactic field have a spin period $P_\textrm{s}$ of less
than 20 ms (called millisecond pulsars;
MSPs)\footnote{http://astro.phys.wvu.edu/GalacticMSPs/GalacticMSPs.txt}. A further 200 or more MSPs reside in globular
clusters\footnote{https://www3.mpifr-bonn.mpg.de/staff/pfreire/GCpsr.html}.

Large pulsar surveys are still being conducted because of the wealth
of information that can be extracted from pulsar observations.  Most
notably, the pulsar timing technique consists of measuring the pulse
times of arrival (TOAs) recorded at an observatory. A model for the
pulsar rotation and the signal propagation from the pulsar to the
observatory is then fitted to the TOAs. The discovery and subsequent
monitoring of relativistic binary systems can yield the strongest
constraints on theories of gravity \cite[PSRs J0737$-$3039A/B,
  J0337$+$1715;][]{ksm+21,vcf+20}, on the equation of state for dense
matter \citep[PSRs J0348$+$0432, J0740+6620;][]{afw+13, NICER_EoS}, and
on the physics of the pulsar radio beam \citep[PSR
  J1906$+$0746;][]{dkl+19}.  Discovering new radio pulsars helps
characterise unidentified gamma-ray sources and pulsar contributions
to the Galactic high-energy diffuse emission \citep{thousandfold}.

An ensemble of MSPs spread over the celestial sphere has the potential
to directly detect the stochastic gravitational wave background (GWB)
generated by a population of supermassive black hole binaries
\citep{saz78,det79,fb90}. Pulsar timing arrays (PTAs) in Australia
\citep{krh+20}, Europe \citep{dcl+16}, and North America
\citep{nanograv21} have been established to pursue this effort.
\citet{gammaPTA} recently demonstrated that gamma-ray timing usefully
complements the radio campaigns.  These individual PTAs have also
combined their efforts to form the International PTA (IPTA)
\citep{ipta19}, later joined by the Indian PTA \citep{inpta19} and
Chinese PTA \citep{lee16}. A key way to improve a PTA's sensitivity is
to increase the number of MSPs, $N_{\textrm{MSP}}$, suitable for
inclusion in the data analysis {{because the signal-to-noise ratio (S/N) for
    the detection of the GWB S/N$_{\textrm{GWB}} \propto
    N_{\textrm{MSP}}$}}. Such MSPs are characterised by a timing
precision of $\lesssim 1 \mu$s over several years and decades
\citep{sej+13}.

The high time and frequency resolution of the pulsar survey data has
also enabled the fortuitous discovery of fast radio bursts (FRBs),
which are bright and short-duration bursts of extragalactic origin.
Some FRBs are seen to repeat and some (but not only repeating FRBs)
have been localised to their host galaxy, potentially serving as
cosmological probes \citep[for a review, see e.g.][]{cc19}.


Pulsar surveys are underway {(or were recently conducted)} at almost
every major radio facility and at various observing frequencies. Near
1.4\,GHz, the list includes the High Time Resolution Universe (HTRU)
South \citep{kjs+10} and the SUrvey for Pulsars and Extragalactic
Radio Bursts \citep[SUPERB;][]{kbj+18} with the Parkes radio telescope.
HTRU North \citep{bck+13} is carried out with the Effelsberg radio
telescope. The Pulsar Arecibo L-band Feed Array (P-ALFA) survey \citep{cfl+06,lbh+15} with the Arecibo
radio telescope was the most sensitive large-scale survey of the
Galactic plane until it was superseded by the Five-hundred-meter
Aperture Spherical radio Telescope (FAST) Galactic Plane Pulsar
Snapshot survey \citep{hww+21}. The unfortunate collapse of Arecibo at
the end of 2020 halted the observations with about 30\% of the
observing towards the inner Galaxy left to be done \citep{psf+22}.
More recently, the Max-Planck-Institut für Radioastronomie Galactic
Plane Survey started a 3000h pulsar survey with the MeerKAT radio
telescope (Padmanabh et al., in prep). Other large-scale surveys at
lower frequencies include the Green Bank North Celestial Cap (GBNCC)
pulsar survey near 350\,MHz \citep{slr+14} and the Canadian Hydrogen
Intensity Mapping Experiment (CHIME) FRB collaboration between 400 and 800
MHz \citep{gac+21}.

At the Nan\c cay Radio Telescope (NRT), a pulsar survey of the Galactic
plane was performed at the end of the 90s \citep{frc+97} using the now
decommissioned Navy Berkeley Pulsar Processor. This survey led to the
discovery of two new young pulsars \citep{tpc+11}. The NRT has also
searched for pulsars in \textit{Fermi} Large Area Telescope (LAT)
gamma-ray sources \citep{des09,cgj+09,gfc+12}.  We report here on the
intermediate latitude SPAN512 pulsar survey with the NRT that began in
2012, exploiting the Nançay Ultimate Pulsar Processing Instrument
(NUPPI) backend \citep{dbc+11}.

The plan of the paper is as follows: we describe the observations
setup in Section~\ref{sec:obs} and the processing pipeline in
Section~\ref{sec:processing}. We present our discoveries in
Section~\ref{sec:results}. Sections ~\ref{sec:discussion} and ~\ref{sec:conclusion} present a discussion of the results and our conclusions, respectively.

\section{Observations}
\label{sec:obs}

\subsection{The Nan\c cay radio telescope}

The SPAN512 pulsar survey was carried out at L-Band (defined as the 1
to 2 GHz part of the electromagnetic spectrum) with the NRT, a transit
telescope of Kraus design oriented north-south, between 2012 and 2016. The sky coverage
($72\degr < l < 150\degr$, $3.5\degr < b < 5\degr$ and $79\degr < l <
150\degr$, $-3.5\degr > b > -5\degr$) is delimited towards the inner
Galaxy by the more sensitive P-ALFA survey \citep{cfl+06,lbh+15} and excludes the
Galactic plane covered by the HTRU-North low-latitude survey
\citep{bck+13}.

\begin{figure}
\includegraphics[height=\columnwidth,angle=-90]{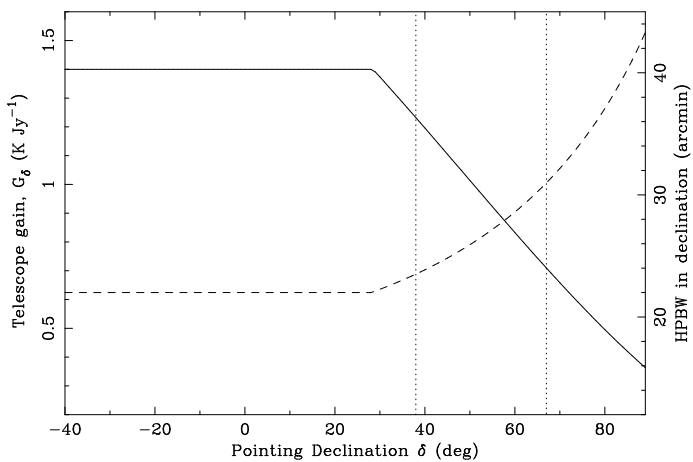}
\caption{Telescope gain (black line) and  HPBW (dashed line) as a function of the NRT pointing Declination for inferior passage. The vertical dotted lines show the Declination range for the sky positions surveyed by SPAN512 with $38^\circ < \delta < 67^\circ$.}
\label{fig:gain}
\end{figure}

The observations use the NRT low-frequency receiver tuned to a central
frequency of 1.466\,GHz, where the system temperature is
$T_{\text{sys}} = 30$\,K, the nominal gain is $G=1.4$\,K\,Jy$^{-1}$
and the half-power beam width (HPBW) is roughly $4'$ (in Right
Ascension $\alpha$) by $22'$ (in Declination $\delta$) {for
  $\delta<25\degr$}. The NRT can observe any source with $\delta >
-39^\circ$. But owing to its specific design, for $\delta > 25\degr$,
the illuminated area of the mirrors, and therefore the gain
$G_\delta$, decreases with higher source declination.
$G_\delta$ can be approximated by
\begin{equation}
G_\delta = 1.5 \times G \times \sin^2 \left(69.13^\circ-\delta/2 \right).
\end{equation}
Conversely, the HPBW increases along the $\delta$ direction, 
\begin{equation}
\text{HPBW}_\delta = 18' / \sin \left(69.13^\circ - \delta/2 \right).
\end{equation}
Figure \ref{fig:gain} illustrates both effects, showing
that the telescope gain is reduced by half at high declinations.
A focal carriage houses the receivers, and moves along a 100m
track, allowing the NRT to follow sources for about one hour (up to two hours for sources with high declination). During source tracking, the
parallactic rotation $\phi$ of the NRT beam is given by
\begin{equation}
\sin(\phi)= \sin(\delta) * \sin(\textrm{HA}),
\end{equation}
where HA is the hour angle of the target on the sky {{(the angle in the east-west direction between the meridian and the source)}}.

\begin{figure*}
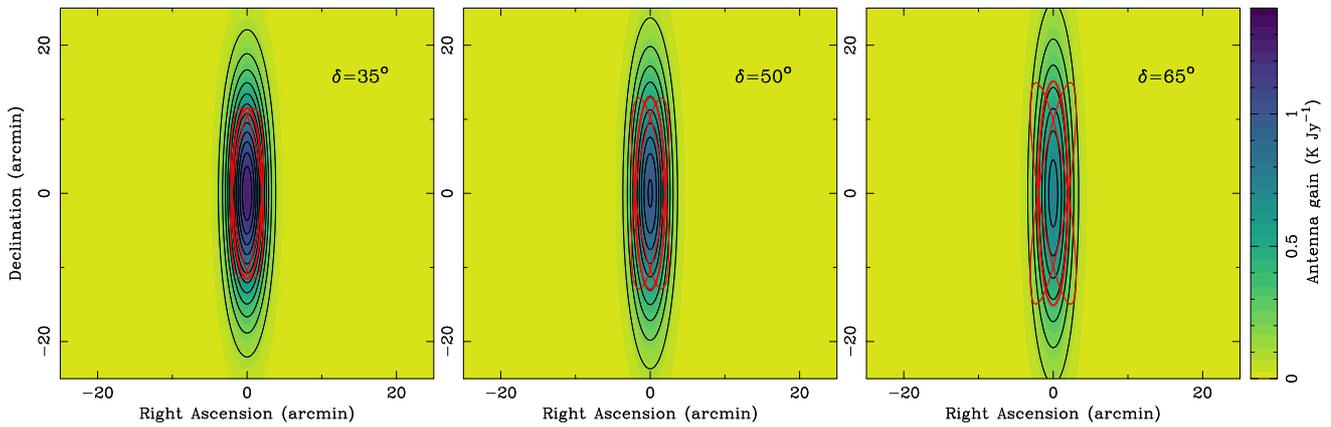

  \includegraphics[width=5.5cm,angle=-90]{beam35.ps}
  \includegraphics[width=5.5cm,angle=-90]{beam50.ps}
  \includegraphics[width=5.5cm,angle=-90]{beam65.ps}
\caption{Beam response in K\,Jy$^{-1}$ of the NRT when pointing sources at three
  different Declinations, $\delta=35^\circ$ (left panel),
  $\delta=50^\circ$ (middle panel), and $\delta=65^\circ$ (right
  panel). In each panel, the black ellipses mark the antenna gain by
  steps of 0.1\,K\,Jy$^{-1}$. The red ellipses represent the rotation
  of the NRT HPBW when observing the sources for HA=-8$^\circ$,
  0$^\circ$ (the meridian) and +8$^\circ$. As the NRT points to
  higher Declination, the beam widens in Declination while the antenna
  gain decreases and the range of the beam rotation angle increases.}
\label{fig:beam}
\end{figure*}

Figure \ref{fig:beam} illustrates the NRT beam pattern and the effect
of parallactic rotation at a frequency of 1.47\,GHz for three
different declinations. The higher the declination, the more elongated
and less sensitive the beam. For high declinations and large hour
angles ($\delta \gtrsim 50^\circ$, $|HA| \gtrsim 10^\circ$), the
parallactic rotation of the elongated beam becomes significant.  To mitigate the
effects of parallactic rotation of the SPAN512 observations and
maintain a more uniform gridding of the sky, we scheduled observations
such that $|HA|<10^\circ$.

Figure \ref{fig:sky} shows the SPAN512 survey sky coverage,
with a total of \NBeam{} pointings. Figure \ref{fig:2215_known} details a small region.
The pointing grid design assumes an idealised rectangular beam with dimensions corresponding to the HPBW in $\alpha$ and $\delta$. The idealised beam assumes observations near the meridian and therefore neglects parallactic beam rotation.
The detailed direction-dependent beam response was studied during the preparation of this article, revealing greater sensitivity variations than intended.
A few known pulsars were therefore not re-detected by the survey, as is discussed in Section $4.1$.
Section $5.1$ uses the true observational grid in comparing our pulsar detections with expectations.

\subsection{Observing setup}

The data are recorded with the versatile NUPPI backend \citep{dbc+11}
based on a CASPER\footnote{https://casper.berkeley.edu}
ROACH\footnote{Reconfigurable Open Architecture Computing Hardware.}
board. For each of the two linear polarisations, a 512\,MHz bandwidth
is first digitised using 8-bit analogue-to-digital converters. Then an
8-tap polyphase filter bank using a Hamming window function
implemented in the Xilinx Virtex 5 field-programmable gate array
(FPGA) splits the band into 1024 channels.  Chunks of contiguous
64\,MHz baseband data are sent to four servers equipped with two
Graphics Processing Units (GPUs) each.

There, the NUPPI software, adapted from the Green Bank Ultimate Pulsar
Processing Instrument (GUPPI) software, is used to detect the signals
from the two linear complex polarisations (i.e. sum the polarisation
to form Stokes I) and to average in time every 64 \us{} before writing
the data in \textsc{PSRFITS} 4-bit search mode format.  The $8\times
64$~MHz sub-bands are later merged together using a modified version of
\textsc{psrfits\_utils}\footnote{https://github.com/demorest/psrfits\_utils}.
The total expected amount of data for the survey was therefore around
53\,TB.


The minimum flux density $S_\textrm{min}$ of a pulsar with spin period $P_\textrm{s}$ detectable by the SPAN512 survey can be written as
{
\begin{equation}
S_\textrm{min} = \frac{\text{S/N} \times (T_{\text{sky}} + T_{\text{sys}})}{G_\delta \sqrt{n_\text{p} t_{\text{obs}} \Delta f}} \sqrt{ \frac{W_\textrm{eff}}{P_\textrm{s}-W_{\textrm{eff}}}}\qquad [\text{mJy}],
\label{eq:radiometer}
\end{equation}
where $\text{S/N}, T_{\text{sky}}, T_{\text{sys}}, n_\text{p},
t_{\text{obs}}, \Delta f, W_{\textrm{eff}} $ denote the pulse
signal-to-noise ratio (S/N), sky and system temperature contributions, the
number of polarisations summed, the duration of the observation, the
observing bandwidth, and the effective pulse width of the  pulsar,
respectively. Here we take $\text{S/N}=8$, $T_{\text{sky}} +
T_{\text{sys}}=35\,$K, $n_\text{p}=2$, $\Delta f=512$\,MHz, and assume
a pulse duty-cycle $W_{\textrm{eff}}/P_\textrm{s}=5\%$}. As this
survey was designed to achieve a sensitivity at the lowest
pointing Declination of  SPAN512 {($\delta\sim38\degr$,
  $G_\delta=1.23$\,K\,Jy$^{-1}$) of $\sim 0.05$ mJy, which is} similar to the
HTRU-North low-latitude survey, $t_\text{obs}$ was set to 18 min. { As
  $t_\text{obs}$ is constant across all pointings, $S_\textrm{min}$
  increases to 0.086\,mJy for the pointings of highest Declination 
  ($\delta \sim 67\degr, G_\delta=0.71$\,K\,Jy$^{-1} $).}

\begin{figure*}
\includegraphics[width=\textwidth]{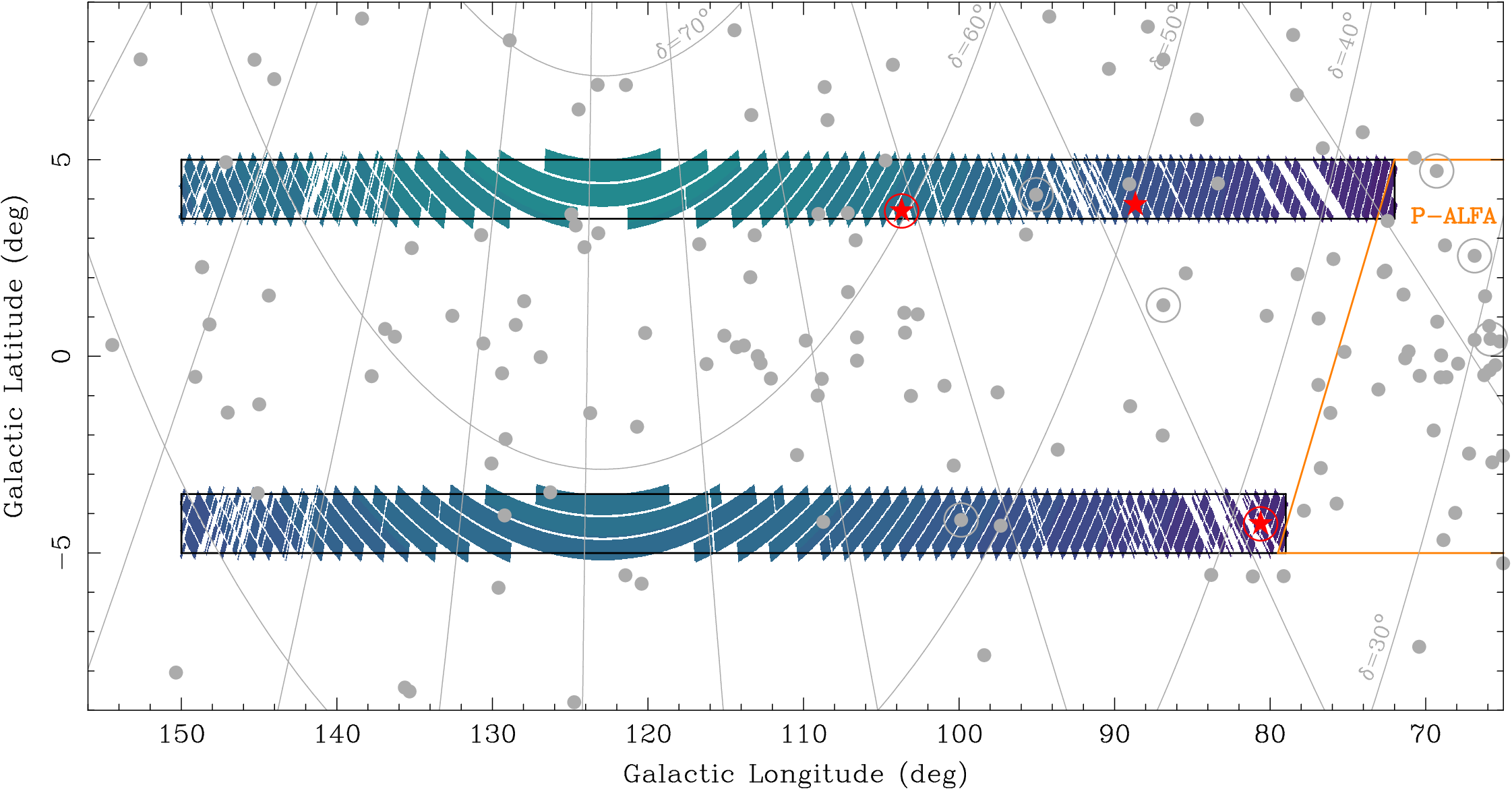}
\caption{SPAN512 survey sky coverage in Galactic coordinates
  shown by the two black rectangles. The same colour scale as in Figure \ref{fig:beam} shows the Declination-dependent survey sensitivity. Grey dots show known pulsars from the
  ATNF Pulsar Catalogue and red stars show the three pulsars
  discovered with SPAN512. Circled symbols indicate MSPs, with spin
  period $< 20$\,ms. The P-ALFA sky coverage is delimited by the
  orange lines.}
\label{fig:sky}
\end{figure*}

Observations started in early 2012 and stopped in June 2016. During
that time, 5412 unique sky pointings were acquired making the survey
89.7\% complete. Table~\ref{tab:survey} summarises the survey
characteristics and Figure~\ref{fig:sky} shows the survey sky
coverage.


\begin{table}
\caption{Summary of the SPAN512 survey parameters.}
\label{tab:survey}
\begin{tabular}{lr}
\hline\hline
Parameter & Value \\ 
\hline
Bandwidth, $\Delta f$ (MHz) & 512 \\
Number of channels, $N_\text{c}$  & 1024 \\ 
Channel bandwidth, $\Delta f_\text{c}$  (MHz) & 0.5 \\
Centre frequency, $f_\text{ctr}$  (MHz) & 1466 \\
Sampling time, $t_\text{samp}$ (\us) & 64 \\
Observation length, $t_\text{obs}$  (min) & 18 \\  
{Sensitivity, $\delta = 38 \degr$ to $67\degr$}, $S_\textrm{min}$ (mJy) & $0.05$ to $0.086$ \\
Sky temperature, $T_\text{sky}$ (K) & 5 \\
System temperature, $T_\text{sys}$ (K) & 30 \\
Sky area (sq. deg) & 224 \\
Number of grid pointings & \NBeam{} \\ 
Number of observed pointings & \NObs{} \\
Completeness of the observations & 89.7\% \\
Completeness of the processing & 100\% \\
\hline
\end{tabular}
\end{table}



\section{Survey processing}
\label{sec:processing}
To analyse the PSRFITS search data, we use the
\textsc{PRESTO}-based \citep{rce03} pipeline developed for processing the P-ALFA survey and described in \citet{lbh+15}. 
Below we detail the processing steps specific to the SPAN512 survey.

\subsection{Radio frequency interferences excision}
The initial processing step is to clean the data to remove
narrow-band radio frequency interference (RFI) with the
\textsc{rfifind} tool from the \textsc{PRESTO} package.
 The median value of the percentage of
data masked is $\sim$ 12\%, with a standard deviation of $4\%$. The 17  pointings with a percentage of masked
data above 30\% were marked for reobservation and 15 of these pointings were reobserved and processed.

\subsection{Dispersion removal}
To detect pulsars with a priori unknown dispersion measure (DM),
that is, the integrated free electron column density along the line of sight,
we mitigate the frequency-dependent dispersive delay by shifting the
frequency channels using a wide range of DM trials. We set the maximum
value of DM to 3000\,pc\,cm$^{-3}$, which is ten times the maximum value
predicted by the \textsc{NE2001} model \citep{cl02} of the Galactic
distribution of free electrons for the different lines of sight of the
SPAN512 survey. By doing this, we can account for some DM excess due to,
for example, the presence of a H$_\textrm{II}$ region on our line of sight and
remain sensitive to high-DM extragalactic fast radio bursts
(FRBs). The time resolution of the data decreases by a power-of-two
downsampling factor when the dispersive smearing inside a single
channel exceeds the sample duration. The spacing between two
neighbouring DM trials, $\delta_{\text{DM}}$, must be limited such
that the dispersive smearing for a DM value halfway between two trial
values is less than the sample duration.  Table~\ref{tab:DD} lists the
de-dispersion plan computed with the \textsc{PRESTO} script
\textsc{DDplan.py} used by the \textsc{prepsubband} program.

\begin{table}
\caption{De-dispersion plan for the SPAN512 survey.}
\label{tab:DD}
\begin{tabular}{rrrrr}
\hline\hline
DM range & $\delta_{\text{DM}}$ & Number of & Downsampling \\ 
(pc cm$^{-3}$) & (pc cm$^{-3}$)   & DM trials & factor \\
\hline
0 - 180 & 0.1 & 1800 & 1 \\
180 - 300 & 0.2 & 600 & 2 \\
300 - 600 & 0.3 & 1000 & 4 \\
600 - 1000 & 0.5 & 800 & 8 \\
1000 - 1800 & 1.0 & 800 & 16 \\
1800 - 3000 & 3.0 & 400 & 32 \\
\hline
\end{tabular}
\end{table}



\subsection{Search}
As for the P-ALFA processing, each de-dispersed time series is then
searched for periodic candidates in two passes using
\textsc{accelsearch}. The first search pass is a zero-acceleration
search summing up to 16 harmonics, aimed at finding isolated pulsars
or pulsars in very wide orbits.  The second pass provides an
acceleration search, with the fundamental harmonic being allowed to
drift by up to 200 Fourier bins. To reduce computing time, the
harmonic sum for this pass is restricted to only eight harmonics. {Each
  search pass produces an independent candidate list.} Only candidates
with $\sigma>2$ probability of not being noise (equivalent Gaussian
significance) are retained in each list \citep{rem02}.

\subsection{Candidate folding}
The candidates from each list are independently sifted to remove those
that are harmonically related, keeping only the ones with highest
significance within groups with similar periods, and having at least
two contiguous DM detections.  After sifting, for up to 150 candidates
with $\sigma>6$ from each list, we form pulse profiles by folding the
raw PSRFITS search data at the candidate period. We find that the
median of the distribution of $\sigma$ values for the 150$^\textrm{th}$ folded
candidate of each pointing is 7, with a median absolute deviation of
0.05.  This confirms that we are unlikely to miss high-significance
candidates due to RFI in the folding stage. In total, we folded approximately $2 \times 800 000$ candidates.

\subsection{Candidate ranking and visualisation}
A set of ratings is applied to all folded candidates, including the
Pulsar Evaluation Algorithm for Candidate Extraction (PEACE) scoring
software \citep{lsj+13} and the Pulsar Image-based Classification
System (PICS) neural-network based artificial intelligence (AI),
trained with P-ALFA candidates \citep{zbm+14}. P-ALFA and SPAN512
candidates have similar observing frequency and bandwidth, and share the
same folding pipeline. Reusing a neural net trained on the  candidates of a different
survey may be suboptimal, but the SPAN512 survey did not
redetect enough known pulsars to retrain it.  The metadata of all candidates 
are stored in a NoSQL MongoDB database for ease in data mining. We
developed
\textsc{psrhunt}\footnote{https://github.com/gdesvignes/psrhunt}, a
Python Qt-based user interface with Matplotlib support \citep{hun07},
to connect to the MongoDB database{ and identify common periodicities
  of radio frequency interferences across different pointings. This
  allowed us to skim more efficiently through the best candidates,}
similar to the \textsc{reaper} \citep{fsk+04} and \textsc{jreaper}
software \citep{kel+09}.

\section{Results}
\label{sec:results}

\subsection{Known pulsars}
Table~\ref{tab:known-psrs} lists 18 pulsars, 15 of which were
discovered within the 224 sq. deg SPAN512 region independently of our
survey. Of these 15 previously known pulsars, 4 are MSPs.
PSR~J2139$+$4716 is a radio-quiet pulsar discovered using {\it Fermi}
LAT data \citep{pletsch_2012}. Two others (PSRs~J0413$+$58 and
J2206$+$6151) lie at unobserved survey grid
positions. Table~\ref{tab:known-psrs} reports the S/N for the
redetection of 6 previously known pulsars and the 6 remaining known pulsars not
redetected by the survey are noted.

Inspection of Figure \ref{fig:sky} shows that all six undetected
pulsars lie between or at the end of beam swaths. We used the Python
notebook \textsc{SPAN512-beams.ipynb} available
online\footnote{https://github.com/gdesvignes/SPAN512} to accurately
estimate the sensitivity of the SPAN512 survey at a given sky
location.

\begin{figure}
\includegraphics[height=\columnwidth,angle=-90]{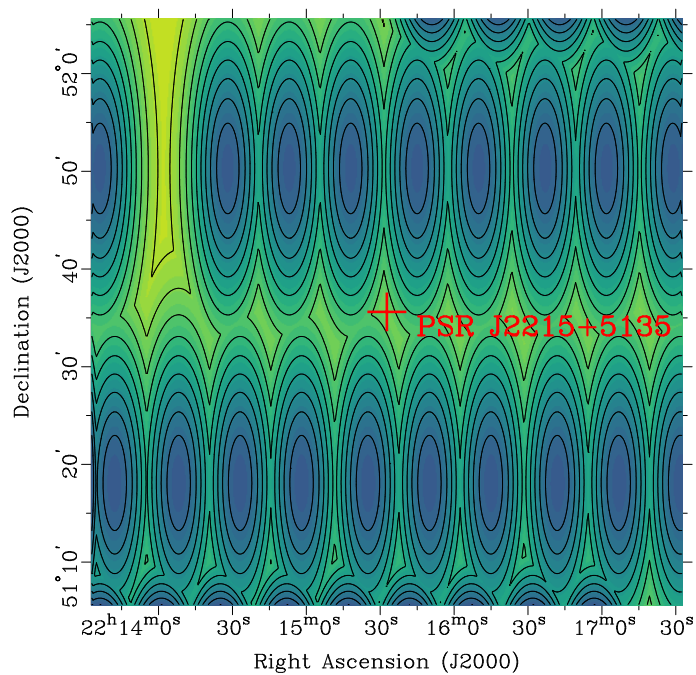}
\caption{Positions of the SPAN512 pointings near the location of the previously known pulsar J2215$+$5135 (depicted with a red cross). The black contours mark the antenna gain by steps of 0.1\,K\,Jy$^{-1}$. The colour scale is the same as in Figure \ref{fig:beam}. This map shows PSR J2215$+$5135 falling between beams with a reduced gain on target of 0.23\,K\,Jy$^{-1}$, explaining its non-detection by the survey. The strip of reduced sensitivity in the top left part of the plot is due to one of approximately $700$ non-observed grid pointings. }
\label{fig:2215_known}
\end{figure}

PSRs J0111$+$6624, J2115$+$5448 and J2215$+$5135 lie in between
observation swathes, with estimated $G$ of respectively 0.32, 0.25, and
0.24 K\,Jy$^{-1}$ and were not redetected.  Figure
\ref{fig:2215_known} illustrates this for PSR J2215$+$5135. Moreover,
PSR J0111$+$6624 has a nulling fraction of $0.33$ \citep[][]{lsk+18}.
For these three sources, we folded the raw PSRFITS data of the closest
pointing using the timing ephemeris available from the ATNF Pulsar
Catalogue, but again, no pulsations were detected.  PSR J2203$+$50 is a
slow pulsar discovered at 350\,MHz with poorly constrained
localisation \citep{hrk+08}.  MSPs J0329$+$50 and J2051$+$50 were
discovered at 820 MHz using the Green Bank telescope, in 30 min
pointings of {\it Fermi} LAT pulsar-like unidentified gamma-ray
sources (Tabassum et al., in prep).  Scintillation also causes  the flux of a
pulsar to vary significantly, hindering detection.

 The six non-detections include all four MSPs. In addition to the two MSP discoveries described below, our pipeline easily detected the fast MSPs J0030$+$0451 and J2317$+$1439 outside the survey area but observed as in the survey grid points. A problem finding fast pulsars is therefore unlikely.
 
 In conclusion, the non-detections are explained by low fluxes, positions where the survey sensitivity is poor, or both. SPAN512 reliably detects active pulsars on the survey grid.
 

\begin{table*}
\caption{List of the 18 currently  known pulsars in the SPAN512 sky area. }
\centering
\begin{tabular}{lrrrrcrrrr}
\hline\hline
PSR Name & \multicolumn{1}{c}{$l$} & \multicolumn{1}{c}{$b$} & \multicolumn{1}{c}{$P_\textrm{s}$} & \multicolumn{1}{c}{DM} & $S_{1.4}$ & Detected & S/N & Remarks\\
         & \multicolumn{1}{c}{(deg)} & \multicolumn{1}{c}{(deg)} & \multicolumn{1}{c}{(ms)} & \multicolumn{1}{c}{(pc cm$^{-3}$)} & (mJy) & & \\
\hline
J0111$+$6624 & 124.93 & 3.61 & 4301 & 111.2 & -- & N  & -- & Between pointings\\ 
J0139$+$5814 & 129.216 & -4.044 & 272 & 73.78 & 4.6 &  Y & 49  & \\
J0329+50 & 146.9 & -4.7 &  3.06 &   7.40 & -- &   N & -- & Poor localisation\\
J0413$+$58 & 147.122 & 4.934 & 687 & 57.0 & 0.19(24)\tablefootmark{a} & -- & -- &  Position not observed\\
J2027$+$4557 & 83.358 & 4.394 & 1099 & 229.59 & 1.34 & Y & 140 &  \\
J2047$+$5029 & 89.057 & 4.376 & 446 & 107.68 & 0.38 &  Y & 10.7 &  \\
{\bf J2048$+$4951} & 88.669 & 3.854 &  568 & 222.8 & 0.12(1) &  Y & 15 & SPAN512 discovery\\
J2051+50 &  89.7 &  4.1 &  1.68 &  61.00 & -- & N  & --  & Poor localisation\\
{\bf J2055$+$3829} & 80.615 &  -4.260 & 2.09 & 91.83 & 0.10 (4)&  Y & 9 & SPAN512 discovery\\
J2115$+$5448 & 95.043 & 4.109 & 2.60 & 77.4 & 0.51(22)\tablefootmark{a} &  N  & -- & Between pointings \\
J2139$+$4716 & 92.633 & -4.02 & 283 & -- & -- &  -- &  -- & Radio quiet\\
J2203$+$50 & 97.322 & -4.307 & 745 & 79.0 & 0.12(15)\tablefootmark{a} & N  & -- &  Poor localisation\\
{\bf J2205$+$6012} & 103.686 & 3.696 & 2.41 & 157.60 & 0.49(1) &  Y  & 14 & SPAN512 discovery\\
J2206$+$6151 & 104.735 & 4.977 & 323 & 167.0 & 0.8 & -- & -- & Position not observed \\
J2215$+$5135 & 99.868 & -4.159 & 2.61 & 69.2 & 0.55(41)\tablefootmark{a} & N & -- &  Between pointings\\
J2229$+$6205 & 107.154 & 3.645 & 443 & 124.61  & 0.8 &  Y & 36 & \\
J2244$+$63 & 109.039 & 3.618 & 461 & 92.0 & -- &  Y & 25 & Poor localisation \\
J2308$+$5547 & 108.729 & -4.206 & 475 & 46.54& 1.9 &  Y & 43 & \\
\hline
\end{tabular}%
\label{tab:known-psrs}
\tablefoot{The three new pulsars discovered with SPAN512 are shown in
  bold.  The columns indicate the pulsar name, the pulsar sky location
  (galactic longitude $l$ and latitude $b$), its spin period
  $P_\textrm{s}$, DM, average flux density at 1.4~GHz ($S_{1.4}$),
  detection status with the SPAN512 survey and the S/N of the folded
  pulse profile produced by \textsc{PRESTO}.
  The parameters for the
  previously known pulsars are taken from the ATNF Pulsar Catalogue
  v1.66.  The parameters for PSR~J2055$+$3829 are from \citet{goc+19}.
  Figures in parentheses represent 1$-\sigma$ uncertainties in the
  last quoted digit.\\
  \tablefoottext{a}{These pulsars fluxes are measured at lower frequencies and scaled
    to 1.4\,GHz assuming the spectral index of $-1.6\pm0.54$ from
    \cite{jsk+17}. For PSRs J0413$+$58,
  J2115$+$5448, J2203$+$50, and J2215$+$5135, we took the flux
  densities and observing frequencies from \citet{scb+19},
  \citet{san16phd}, \citet{scb+19}, and \citet{hessels_2011},
  respectively, as we noted discrepancies with the values
  reported in the ATNF Pulsar Catalogue.}
  }
\end{table*}



\subsection{New detections}
In addition to nine detections of previously known
pulsars, the SPAN512 survey discovered two new MSPs PSR J2055$+$3829 and PSR J2205$+$6012, and the slow pulsar \psrb{}. \citet{goc+19} described the discovery and timing of PSR J2055$+$3829. We report the remaining two here.  

Follow-up NRT observations were recorded with the NUPPI backend in
regular timing observing mode \citep[see e.g.][]{goc+19}.  Data postprocessing uses the \textsc{PSRCHIVE} pulsar software \citep{hvm04,vdo12}. The data are calibrated in polarisation using polarised noise diode data obtained preceding each observation. Regular observations
of the quasar 3C286 and the radio galaxy 3C123 are also used to flux-calibrate the data.

\subsubsection{PSR J2048$+$4951}
\psrb{} is a slow pulsar with $P_\textrm{s}=568$\,ms and
$\textrm{DM}=223$~pc~cm$^{-3}$. 
Between 2013 and 2019, we performed 56 follow-up observations at L-band with the NRT.  The TOAs of the daily-averaged and frequency-summed pulse profiles were modelled
with the \textsc{Tempo2} \citep{hobbs2006} timing software and the Jet Propulsion Laboratory (JPL) DE438 Solar
System ephemeris. The timing residuals are shown in
Figure \ref{fig:2048_resid} and the best-fit parameters are reported in
Table~\ref{tab:param2048}.  We used this ephemeris to integrate all
calibrated pulse profiles and build the high-S/N profile shown in
Figure \ref{fig:2048_profile}. From this high-S/N pulse profile, we
determine the rotation measure (RM) ---the frequency-dependent rotation of the position angle of the linear polarisation of the radio waves due to propagation in the magnetised interstellar medium--- to be  $\textrm{RM}=-196(10)$\,rad\,m$^{-2}$. The mean flux density of the average pulse profile at 1.4~GHz is $S_{1.4}=0.12(1)$ mJy. 

\begin{figure}
\includegraphics[height=\columnwidth,angle=-90]{2048_residuals.ps}
\caption{Timing residuals of \psrb{}.}
\label{fig:2048_resid}
\end{figure}

\begin{figure}
\includegraphics[height=\columnwidth,angle=-90]{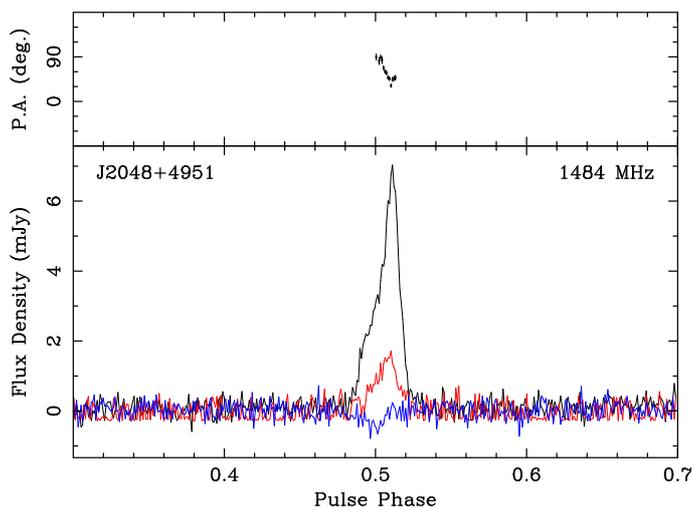}
\caption{Polarisation profile of \psrb{}. Lower panel: Black, red, and blue lines represent the total intensity, linear, and circular
  polarisation, respectively. The upper panel shows the linear polarisation position angle. }
\label{fig:2048_profile}
\end{figure}

\begin{table}
\caption{Parameters for PSR J2048$+$4951.}
\label{tab:param2048}
\begin{tabular}{ll}
\hline\hline
PSR Name & J2048$+$4951\\ 
\hline
MJD range & 56434 -- 58548 \\ 
 Number of TOAs & 56 \\ 
 rms timing residuals ($\mu s$) & 329 \\ 
 Reference epoch (MJD) & 56414 \\ 
\\ 
Measured timing parameters\\ 
\\ 
Right ascension, $\alpha$ (hms, J2000) & 20:48:55.982(4)  \\ 
Declination, $\delta$ (hms, J2000) &  +49:51:53.02(3) \\ 
Spin frequency, $\nu$ (s$^{-1}$) & 1.759630876491(16) \\ 
$\dot{\nu}$ (s$^{-2}$) & $-$1.63756(5)$\times 10^{-14}$  \\ 
$\ddot{\nu}$ (s$^{-3}$) & 1.59(6)$\times 10^{-25}$  \\ 
DM (cm$^{-3}\,$pc) & 222.8 \\ 
\\ 
Derived parameters\\ 
\\ 
Characteristic age, $\tau_c$  (Myr) & 1.7 \\
Spindown power, $\dot E$ (erg s$^{-1}$) & $1.14\times 10^{33}$ \\
Distance (NE2001, kpc) & 9.2\\
Rotation Measure, RM (rad\,m$^{-2}$) & -196(10)\\
$S_{1.4}$ (mJy) & 0.12(1) \\ 
\hline
 \end{tabular}
\tablefoot{We used the JPL DE438 Solar System ephemeris and the
  Barycentric Coordinate Time (TCB) timescale in our \textsc{Tem\ po2}
  timing analysis. The values in parentheses represent the 1-$\sigma$
  error bars reported by \textsc{Tempo2}.}
\end{table}



\subsubsection{PSR J2205$+$6012}
The SPAN512 survey discovered the millisecond pulsar J2205$+$6012 in
2013, with $P_\textrm{s}=2.41$\,ms and $\textrm{DM} \sim
157~\textrm{pc}~\textrm{cm}^{-3}$, in a 1.1-day orbit. Since then, it
has been regularly monitored at the NRT with its L-band
receiver. Since mid-2019, PSR J2205$+$6012 has been observed monthly
with the Effelsberg radio telescope at S-band (defined as 2 to 4 GHz)
with a few observations at C-band (4 to 8 GHz). Due to the low number
of C-band observations, these data were not included in the timing
analysis but were used to constrain the pulsar spectrum.  All
Effelsberg data are recorded with the PSRIX backend and calibrated in
polarisation and flux using observations of a pulsed noise diode and
the planetary nebula NGC~7027. More information about the PSRIX system
is provided by \citet{laz16}.

With only the last two years of our dataset providing multifrequency
coverage at S-band, we split the large fractional bandwidth of the
NUPPI backend into four subbands. The Effelsberg data are fully summed
in frequency.  Templates for the L-band and S-band averaged pulse
profile are built from de-noised time-integrated pulse profiles with
the wavelet noise removal software \textsc{psrsmooth} from
\textsc{PSRCHIVE}.  Before determining the TOAs through
cross-correlation of the pulse profiles with the high-S/N templates,
the NRT daily pulse profiles are fully integrated in time (from 30 min
to 1.5 h) while the Effelsberg profiles are integrated for up to 35
min.  In total, we have 852 and 47 TOAs at L-band and S-band,
respectively. Following \citet{dcl+16}, we used \textsc{TEMPONEST}
\citep{lah+14} to perform the timing analysis and explore the
parameter space of the pulsar timing model. Beyond the astrometric and
spin parameters, the timing model includes the `ELL1' parametrisation
\citep{lcw+01} of the pulsar motion around the centre of mass. A set
of `error scaling factor' EFAC and `error added in quadrature' EQUAD
parameters are required to properly weight the estimated TOA
uncertainty between each observing system. The EFAC and EQUAD
parameters are sampled with uniform and log-uniform priors in the
log$_{10}$- range $[-0.5, 1.5]$, $[-10, -3],$ respectively. The model
also includes DM variations and red timing noise (TN) described as
stationary stochastic processes, with power-law spectra described as
$\frac{A^2}{12\pi^2}\left(\frac{f}{f_r}\right)^{-\gamma}$, a
dimensionless amplitude $A$ at reference frequency $f_r$ of 1
yr$^{-1}$ and the spectral index $\gamma$. $A$ and $\gamma$ are
respectively sampled from a log-uniform prior in the log$_{10}$ range
[-20,-8] and a uniform prior in the range [0,7], for both DM and TN
modelling. See also \citet{cll+16} for more details.
\textsc{MultiNest} \citep{fhb09} sampling uses 2000 live points.

\begin{figure*}
\includegraphics[width=\textwidth]{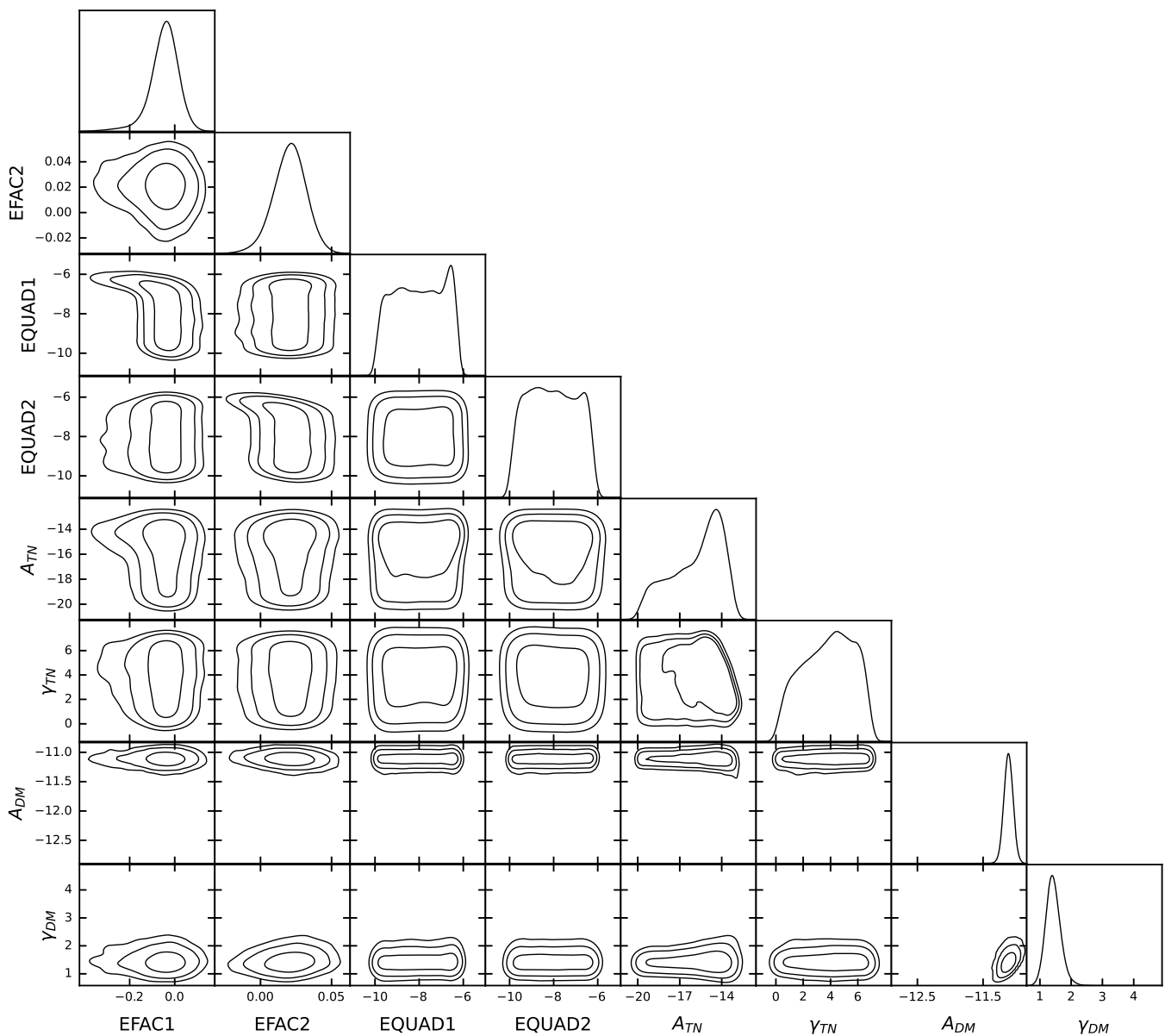}
\caption{Triangular plot of the posterior distributions for the noise
  parameters included in the \textsc{TempoNest} analysis of PSR
  J2205+6012. The parameters EFAC1, EQUAD1, and EFAC2 on the one hand and EQUAD2 on the other shown in
  log-space refer to the Effelsberg and NRT data, respectively. The
  last four parameters are the amplitude and spectral indices for the
  TN ($A_{\textrm{TN}}$, $\gamma_{\textrm{TN}}$) and DM
  ($A_{\textrm{DM}}$, $\gamma_{\textrm{DM}}$),
  respectively. $A_{\textrm{TN}}$ and $A_{\textrm{DM}}$ are also shown
  in log-space. The set of EFAC and EQUAD parameters shows no evidence
  of systematic errors in our TOA uncertainties. While the DM noise is well
  constrained, the TN remains unconstrained with our eight-year dataset.}
\label{fig:triangle}
\end{figure*}

Table~\ref{tab:2205} lists the timing parameters derived from the
\textsc{TempoNest} posterior distributions. A
Tempo\footnote{http://tempo.sourceforge.net/} ephemeris is also
available online for monitoring by other radio telescopes (see footnote
8). With our current dataset, the TN remains unconstrained (see
Fig.\,\ref{fig:triangle}) and we can set the 95\% confidence upper
limit on its amplitude $A_{\textrm{TN}}$ with $\log_{10}(
A_{\textrm{TN}}^{95\%}) < -13.2$. The DM noise is well constrained and
we can reconstruct the DM waveform showing variations of about $ 0.02$
pc cm$^{-3}$ during the last 8 years.  Figure \ref{fig:2205_timing}
presents the timing residuals using the maximum likelihood timing
parameters and after the DM variations waveform has been subtracted
from the residuals. These residuals are characterised by a weighted
root mean square (rms) of 1.5 $\mu$s. Following \citet{nanograv15a},
we also computed the weighted rms after averaging the NRT residuals
from the four frequency sub-bands of the same epoch and found a rms of
752\,ns.

 We find no parallax signature, which is unsurprising because for the high
 ecliptic latitude of the  MSP of $63^\circ$, the annual amplitude is $250/d$ ns for a pulsar distance $d$ in kiloparsecs. Both NE2001 and the newest YMW16
 Galactic electron density model \citep{ymw16} predict
 $d>3$\,kpc. Post-Keplerian parameters such as the Shapiro delay or
 the orbital period derivative are not detectable with the current
 dataset and so the masses of the two stars are not accurately
 measured. The mass function
\begin{equation}
f(m_\textrm{p},m_\textrm{c}) = \frac{4\pi^2 x^3}{P_\textrm{b}^2 T_\odot} = \frac{(m_\textrm{c} \sin{i})^3}{(m_\textrm{p}+m_\textrm{c})^2} = 0.001487903104 \,M_\odot, 
\end{equation}
where $T_\odot=GM_\odot/c^3=4.925\, \mu$s gives a lower limit for
the companion mass $m_\textrm{c}=0.14$\,$M_\odot$, assuming a low
pulsar mass $m_\textrm{p}$ of 1.2~$M_\odot$ and an edge-on orbit
(inclination angle $i=90^\circ$). The projected semi-major axis $x$
and the orbital period $P_b$ are given by the timing analysis. This
constraint argues for a low-mass helium white dwarf as the companion
star of PSR~J2205$+$6012 {(see e.g. \cite{tau11} for a review of the
  evolutionary path of this type of binary)}. No optical counterpart
from the white dwarf was found in the Digitized Sky Survey at the
position of the pulsar.

\begin{figure*}
\centering
\includegraphics[height=\textwidth,angle=-90]{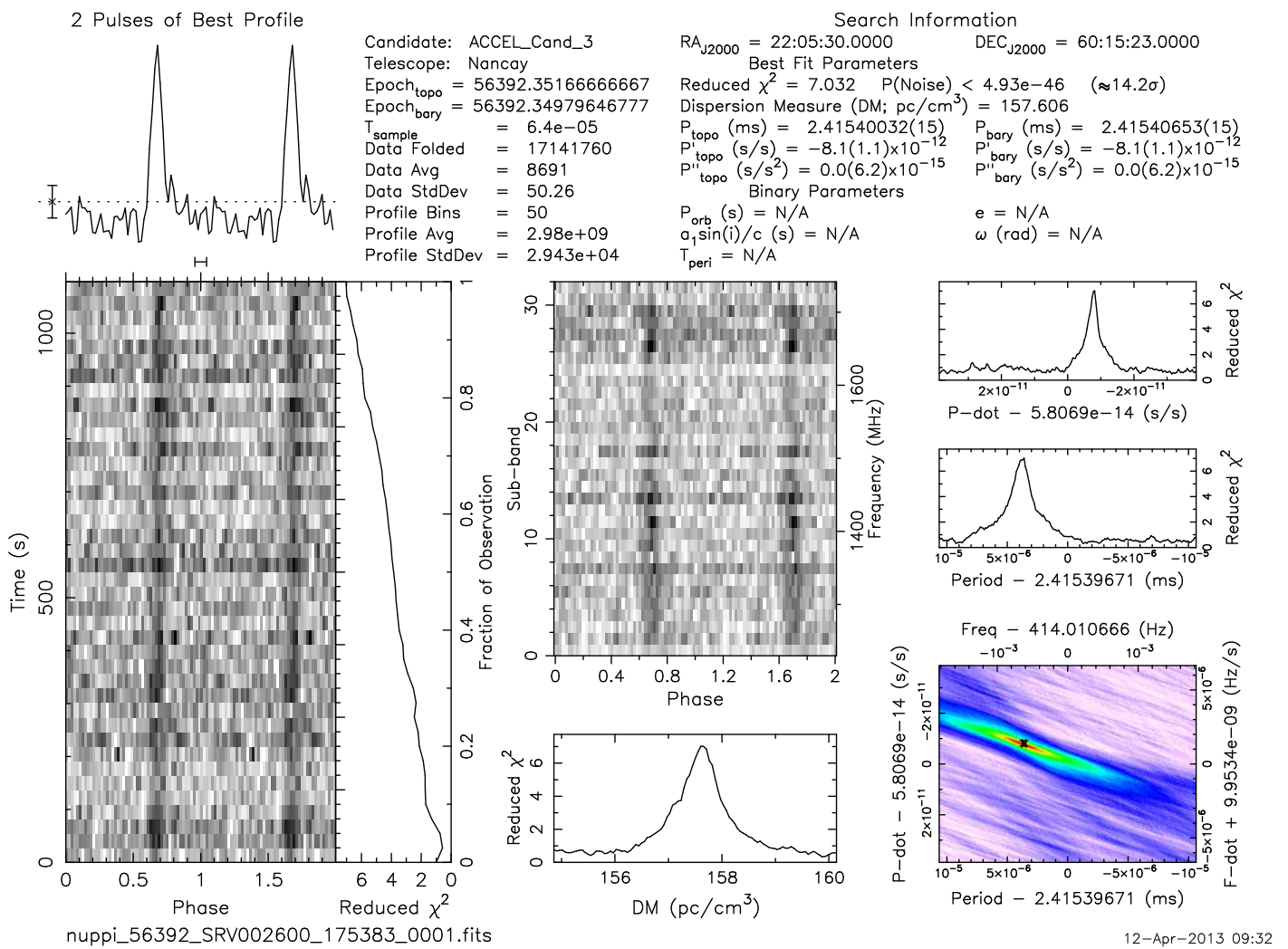}
\caption{\small Discovery plot of PSR~J2205$+$6012 produced by \textsc{PRESTO}.\label{fig:2205_discovery}}
\end{figure*}

\begin{figure}
\includegraphics[height=\columnwidth,angle=-90]{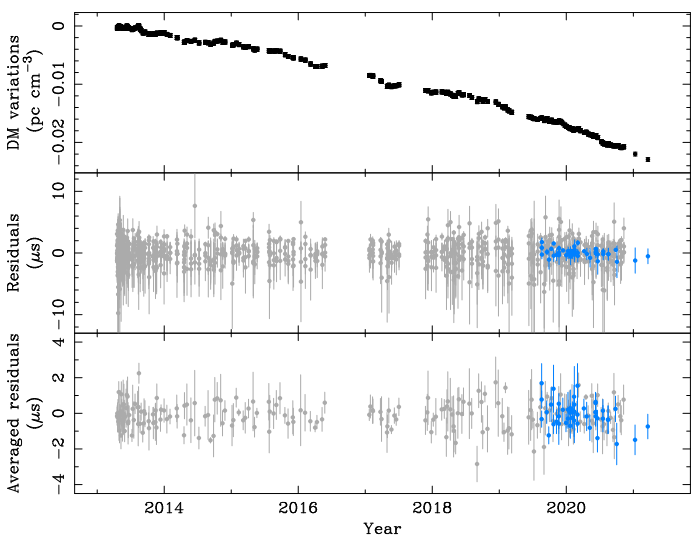}
\caption{Timing results of \psra{}. Top panel: Waveform of the DM
  variations from the maximum likelihood results of the TempoNest
  analysis. Middle panel: Timing residuals after
  subtraction of the ML DM variations waveform. Bottom panel: Same as
  middle panel but with the averaged timing residuals from the NRT
  subbanded data. Grey points and blue triangles represent the NRT
  L-band and Effelsberg S-band data, respectively.}
\label{fig:2205_timing}
\end{figure}

\begin{figure}
\includegraphics[height=\columnwidth,angle=-90]{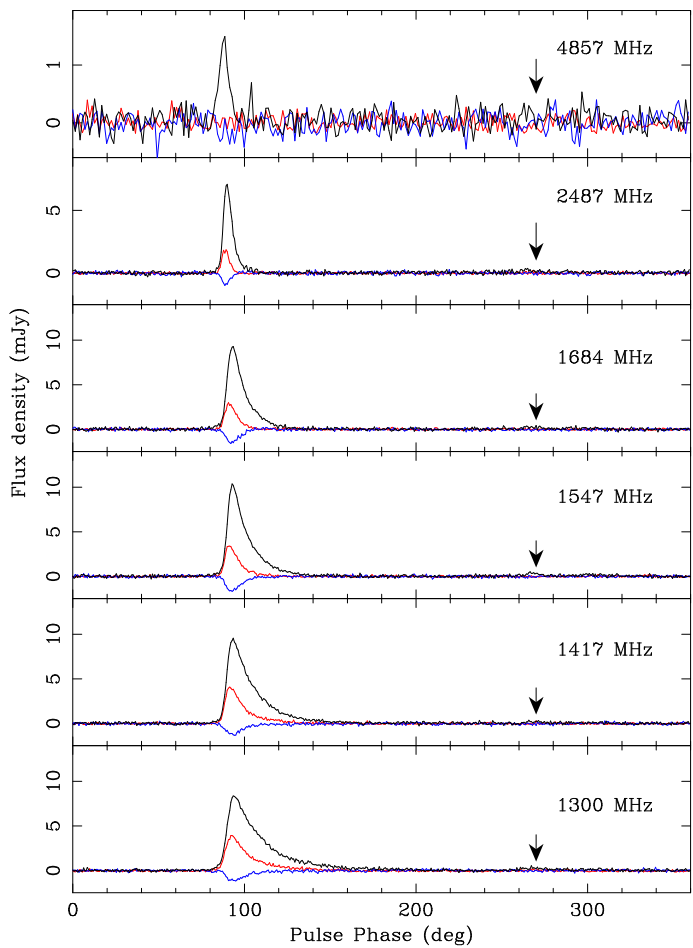}
\caption{Polarisation profiles of \psra{} for decreasing observing
  frequencies. The black, red, and blue lines represent the total
  intensity, linear, and circular polarisation, respectively. The arrows indicate the pulse phase for the second weak component of the pulse profile.}
\label{fig:2205_profiles}
\end{figure}

The discovery plot (see Fig. \ref{fig:2205_discovery}) and the data
from the various frequency bands shown in Fig.
\ref{fig:2205_profiles} show a profile with a single scattered
component. Integrating the L-band data, as in Fig.\,\ref{Gprofile},
reveals another weak component separated by half a rotation. Because
of the lower S/N of the S-band data, there is only a tentative
detection of this second component at this frequency.  This suggests
that PSR J2205$+$6012 might be an orthogonal rotator. Unfortunately,
the L-band profile is quite scatter broadened and the S-band profile
shows little linear polarisation, with a narrow duty cycle. Assuming a
dipolar magnetic field, the Rotating Vector Model \citep[RVM;][]{rc69}
describes the position angle (PA) of the linear polarisation as the
projection of the magnetic field lines rotates with the magnetic axis
around the spin axis of the pulsar. Unfortunately, fitting the RVM to
the PA does not reliably constrain the magnetic inclination
angle or viewing geometry of the  pulsar and we are not able to confirm whether or not
PSR~J2205$+$6012 is indeed an orthogonal rotator.

The pulse width at the 50\% peak intensity, $W_{50}=35~\mu$s, is
significantly narrower than that of the precisely timed pulsar
J1909$-$3744 \citep{jbv+03}. From the S-band, C-band, and sub-banded
L-band data, we estimate the spectral index to be -2.2(4), which is consistent
with the value for the MSP population \citep{kxl+98}.  Assuming an
exponential scattering tail and an intrinsic Gaussian pulse profile,
we applied the \textsc{scatternest}
tool\footnote{https://github.com/gdesvignes/scattering}
to the total averaged pulse profile
reduced to eight
sub-channels to estimate the scatter parameters. We find the scattering
time at 1~GHz, $\tau_{\rm 1\,GHz}=0.32\pm0.01$\,ms, the scatter index $\alpha=-3.71\pm0.06,$
and $\textrm{DM}=157.643\pm0.001$\,pc\,cm$^{-3}$.

\begin{table*}
\caption{Parameters for \psra.}
\begin{minipage}{180mm}
\label{tab:2205}
\begin{tabular}{ll}
\hline\hline
PSR Name &  J2205$+$6012 \\ 
\hline
MJD range & 56400 -- 59294   \\ 
 Number of TOAs &  899 \\ 
 rms residuals ($\mu$s) & 1.50 \\ 
 rms averaged residuals ($\mu$s) & 0.752 \\ 
 {TOA median uncertainty at 2.5 GHz, $\sigma_{\textrm{TOA}_{2.5}}$ ($\mu$s)} & 0.61\\
 Reference epoch (MJD) &  56400 \\ 
\\ 
Measured timing parameters\\ 
\\ 
Right ascension, $\alpha$ (hms, J2000) &  22:05:34.201416(20) \\ 
Declination, $\delta$ (dms, J2000) &  +60:12:55.14068(14) \\ 
Proper motion in $\alpha$, $\mu_\alpha \cos \delta$  (mas\,yr$^{-1}$) & -4.318(27)  \\ 
Proper motion in $\delta$, $\mu_\delta$  (mas\,yr$^{-1}$) & -3.082(24)  \\ 
Period,  $P_\textrm{s}$ (ms) &  2.41554794898110(4)\\ 
Period derivative, $\dot{P_\textrm{s}}$ ($\times 10^{-20}$) &  1.979391(29)\\ 
DM (cm$^{-3}$pc) & 157.6435(6)  \\ 
DM derivative, ${\textrm{DM1}}$ (cm$^{-3}$pc yr$^{-1}$) & -0.0015(4) \\
DM second derivative, ${\textrm{DM2}}$ (cm$^{-3}$pc yr$^{-2}$) & -0.00017(6)\\ 
Orbital period, $P_b$ (d) & 1.094551337704(12) \\ 
Projected semi-major axis, $x$ (lt-s) & 1.18410030(8) \\ 
ELL1 Laplace-Lagrange parameter $\kappa$ & $8.2(1.3) \times 10^{-7}$ \\ 
ELL1 Laplace-Lagrange parameter $\eta$ & $-0.5(1.4) \times 10^{-7}$ \\ 
Time of ascending node (MJD) &  56306.555782949(22) \\ 
$\log_{10}(A_\textrm{DM})$ & $-11.12_{-0.06}^{+0.07}$\\
$\gamma_\textrm{DM}$ & $1.44_{-0.26}^{+0.20}$\\
$\log_{10}(A_\textrm{TN}^{95\%})$ & $<-13.2$\\
$\gamma_\textrm{TN}$ & ---\\
\\ 
Derived parameters\\ 
\\ 
Gal. longitude, $l$ (deg) & 103.7   \\ 
Gal. latitude, $b$ (deg) & 3.7   \\
Radio flux at 1.30 GHz (mJy) & 0.55 \\
Radio flux at 1.42 GHz (mJy) & 0.49 \\
Radio flux at 1.55 GHz (mJy) & 0.42 \\
Radio flux at 1.68 GHz (mJy) & 0.34 \\
Radio flux at 2.50 GHz (mJy) & 0.14 \\
Radio flux at 4.85 GHz (mJy) & 0.03 \\
Flux density spectrum & -2.2(4) \\
Rotation measure, RM (rad m$^{-2}$) & -85(2) \\
Scattering time at 1GHz (ms) &  0.32(1)\\
Scattering index & -3.71(6)\\
Pulse width at 2.55 GHz ($\mu s$) & 35\\

Spindown power $\dot E$ (erg s$^{-1}$) & $6.37 \times 10^{34}$\\
Integral energy flux $>100$ MeV, $G_{100}$ (erg cm$^{-2}$ s$^{-1}$) &  3.7(1.0)$\times 10^{-12}$ \\ 
Distance (YMW16, kpc)  & 3.54\\ 
Distance (NE2001, kpc) & 5.50\\ 
Luminosity $L_\gamma$ (YMW16 distance,  erg s$^{-1}$) & $5.6(1.6)\times 10^{33}$  \\ 
Luminosity $L_\gamma$ (NE2001 distance, erg s$^{-1}$) & $13.6(3.8)\times 10^{33}$  \\ 
Efficiency $\eta = L_\gamma / \dot E$ (YMW16 distance, percent)  &  8.8 \\
Efficiency $\eta = L_\gamma / \dot E$ (NE2001 distance, percent) &  21.  \\

Total proper motion, $\mu$ (mas\,yr$^{-1}$) & 5.336(15) \\ 
Characteristic age, $\tau_c$ (Gyr) &  1.93 \\ 
Surface magnetic field, $B$ ($\times 10^8$ G) &  2.2\\ 
 & \\
\hline
\end{tabular}
\tablefoot{Our \textsc{TEMPONEST} timing analysis makes use of the JPL
  DE438 Solar System ephemeris and the TCB timescale. For the measured
  timing parameters, the numbers in parentheses represent the
  1-$\sigma$ uncertainty taken as the 68\% confidence levels around
  the median value for the posterior of each parameter. As the TN is not
  constrained, we report the 95\% confidence upper limit for the
  amplitude, $A_{\textrm{TN}}^{95\%}$. {The rms residuals are computed
    after subtraction of the DM model.}}
\end{minipage}
\end{table*}



\subsubsection{Gamma-ray pulsations}
\label{GammaSection}
{In the $10^{34} \leq \dot E < 10^{35}$ erg s$^{-1}$ decade where PSR
  J2205+6012 lies, with $\dot E = 4\pi^2 I_0 \dot{P_\textrm{s}}/P_\textrm{s}^3$ being the
  spindown power (assuming a moment of inertia for the neutron star of 
  $I_0=10^{45}$\,g\,cm$^{-2}$), over 70\% of MSPs show gamma-ray
  pulsations in {\it Fermi} LAT data \citep{thousandfold}. In
  comparison, $\sim 30$\% of MSPs in the next lower decade (like PSR
  J2055+3829, $\dot E = 4.3 \times 10^{33}$ erg s$^{-1}$) pulse in
  gamma rays.  Only a few percent of middle-aged pulsars in the
  $10^{33}$ decade, like PSR J2048+4951, are seen by the LAT.}

\begin{figure}
\centering
\includegraphics[width=\columnwidth]{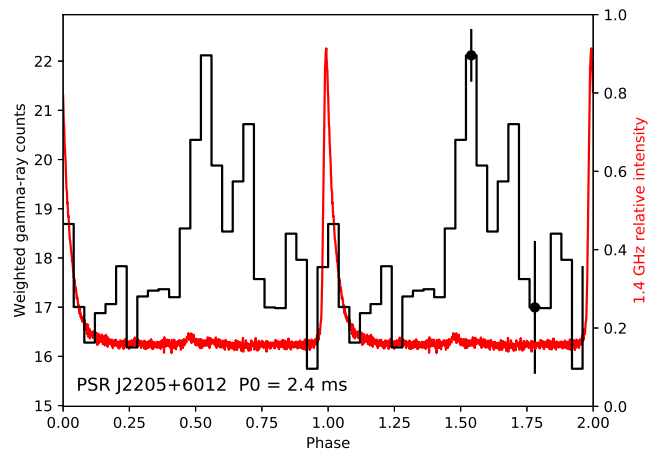}
\caption{\small  Top frame: Weighted pulse profile for $>100$ MeV gamma-rays within $2^\circ$ of the  direction to PSR J2205+6012, 
recorded between MJD 54682 and 59354 (25 bins). 
The largest and smallest uncertainties are shown. 
The overlaid phase-aligned Nan\c cay radio profile is the sum of the four lowest frequency bands shown in Figure \ref{fig:2205_profiles}, allowing the faint peak at phase $0.5$ to appear.
Two rotations are shown for clarity.
\label{Gprofile}}
\end{figure}

PSR J2055+3829 shows no hint of gamma-ray emission \citep{goc+19}.  To
search for gamma-ray pulsations from PSRs J2048+4951 and J2205+6012, we
selected gamma-ray photons within $2^{\circ}$ of the pulsar position,
near the energy-dependent point-spread-function of the  LAT for our
$E_\gamma > 100$ MeV data sample.  We converted their arrival times in
the LAT to neutron star rotational phases using the ephemeris
parameters in Tables \ref{tab:param2048} and \ref{tab:2205} and the
{\tt fermi} plug-in \citep{Ray2011} to {\textsc Tempo2}.  Weighting
the photons according to their angular distance from the pulsar and
their energy optimises the S/N in the presence of background from
nearby sources and diffuse emission.  We used the `simple' method
described by \citet{SearchPulsation}.  The $\mu_w$ parameter of this latter method
gauges pulsar spectral hardness relative to that of the local
background. \citet{thousandfold} found that values $\mu_w= (3.0,\,
3.6, \,4.2)$ cover the range observed to maximise signal significance
for a large sample of gamma-ray pulsars, while maintaining a small
number of trials. Smaller (larger) $\mu_w$ values correspond to softer
(harder) pulsar spectra. PSR J2048+4951 shows no hint of gamma-ray
emission, but PSR J2205+6012 is clearly detected, as in Figure
\ref{Gprofile}.  A weighting parameter value $\mu_w = 4.2$ maximises
the weighted H-test value to 57, for $6.4\sigma$ pulsed significance
\citep{DeJager2010, KerrWeighted}, without trial corrections.

Figure \ref{Gprofile} also shows the phase-aligned $1.4$-GHz radio
profile. Phase $0$ is set to the peak in the radio profile, and the
possible error of the gamma-ray alignment due to the uncertainty in
the  DM is $<0.01$ in phase.  The gamma-ray peak
is roughly aligned with the small radio peak near $0.5$ in phase,
which is typical of the aligned MSP category defined by \citet{egc+12}.

Table \ref{tab:2205} includes some gamma-ray properties.  The integral
energy flux above 100 MeV, $G_{100}$, comes from 4FGL-DR3, the {\it Fermi}
LAT fourth source catalogue, using 12 years of data \citep{4FGL-DR3}. In 4FGL,
the source is too faint ($4.5\sigma$ significance) to fit a curved
spectral shape.  However, the spectral points extend beyond 10 GeV, and
$\mu_w = 4.2$ suggests that the spectrum is hard, with a
cut-off energy at the high end of the range shown in Figure 7 of the
{\it Fermi} LAT pulsar catalogue, 2PC \citep{2PC}.  On the other
hand, the gamma-ray
efficiency $0.09 < \eta = L_\gamma / \dot E < 0.21$ is typical for a millisecond pulsar (see 2PC Figure 10), albeit
dominated by the large uncertainties for the distance $d$ and the
beaming fraction $f_\Omega$ (defined in 2PC Equation 16, and used in
$ L_\gamma = 4\pi d^2 f_\Omega G_{100}$).

\subsection{Single pulse and FRB search}
All SPAN512 pointings were searched for radio transients using the
\textsc{PRESTO} \textsc{single\_pulse\_search.py}, which is similar to the
P-ALFA pipeline as described in \cite{lbh+15}.
This search applied
matched-filters with boxcars up to 0.1s, retaining only pulses with
S/N>6. Assuming the single pulse radiometer equation \citep{mc03},
this translates into a fluence of 0.2 to 1\,Jy\,ms for single pulse
widths of 1 to 25\,ms, respectively. No FRBs were detected.  Even
assuming the upper range value for an FRB rate of $10^3$ to $10^4$
sky$^{-1}$ day$^{-1}$ with fluence $>1$ Jy\,ms \citep{cc19}, our
current sky coverage of $\sim$200 sq. deg. with 18 min pointings gives
$<1$ FRB event potentially detectable by our survey, which is consistent with
our actual lack of detection.

\section{Discussion}
\label{sec:discussion}

\subsection{Expected yield}
We estimated the Galactic population of non-recycled pulsars covered
by the SPAN512 sky area using the PsrPopPy package
\citep{blr+14}. Following \citet{lfl+06} and \citet{fk06}, we assume
log-normal distributions to describe the pulsar periods $P$ and
luminosities $L$ with $< \log P > =2.7$, $\sigma_{\log P}=-0.34$ and
$< \log L >=-1.1$, $\sigma_{\log L}=0.9$, respectively. The simulated
pulsar populations follow a radial distribution with a normal
distribution around the plane defined by a scale height of 0.33 kpc
\citep{lfl+06}.  We ran 500 simulations with a population large enough
for 1121 pulsars to be detected by the PMPS. 
{For each simulation, PsrPopPy uses the Declination-dependent beam shape of the NRT and the grid of observed SPAN512 pointings to compute the number of pulsars detectable with a S/N threshold of 8.
We find that the SPAN512 survey should detect between 9 and 25 non-recycled pulsars (taken as the 95\% probability contours). }

{Our detection (discovery) of 7 (1) young non-recycled pulsars is below the predicted
range.}
{One reason may be that the predictions are sensitive to the $P$, $L,$ and scale-height parameters, which may overestimate the pulsar population in the survey region.}
{Another cause described by \citet{lbh+15} is the presence of RFI and red noise in the data that degrades sensitivity towards slow pulsars compared to the theoretical prediction from the radiometer equation (Eq.~\ref{eq:radiometer}). 
Similar issues may have caused our pipeline to miss some pulsars.}

For MSPs, an adaptation of the simple population synthesis of \citet{JohnstonSmith2020} using parameter distributions similar to those of \citet{Levin2013HTRU_VIII} predicts about two MSPs for the survey zone, one of which would be detectable in gamma rays\footnote{Simon Johnston, private communication.}. {We did indeed detect two MSPs, both new discoveries, with gamma rays from one of them. Four other known MSPs are in the region, but are undetected by the survey, as discussed above; half of these are detected in  gamma rays.}

\subsection{Distance to PSR~J2205$+$6012}
With a DM $= 157.6$~pc~cm$^{-3}$ for PSR~J2205$+$6012, the NE2001 model predicts a distance of 5.5 kpc, placing the pulsar between the Perseus and Outer arms. NE2001 also predicts a scattering time scale of 0.013 ms at 1 GHz for this line of sight, which is 25 times less than the observed value. \citet{ymw16} predict a smaller distance of 3.54 kpc, placing the pulsar within the Perseus arm. 

PSR~J2205$+$6012 lies 0.45 degrees from Cep 15, a B1-type star at a distance of 810 pc, for a projected distance of 6.4 pc between the star and the line of sight to the pulsar. Cep 15 is hot enough to have a Str\"omgren sphere \citep{str39} of tens of parsecs of ionised material that could contribute to the observed DM and scattering.
Subtracting{ an extra contribution of e.g. 30 ~pc~cm$^{-3}$ due to the Str\"omgren sphere} from the DM, the NE2001 and YMW16 distances become $4.4$ and $3.2$ kpc, respectively, and the gamma-ray luminosity $L_\gamma$ remains in the typical range. On the other hand, if such a large fraction of the electron column density along the line of sight was concentrated near Cep 15, it might explain the large scattering tail.

\section{Conclusions}
\label{sec:conclusion}
We present SPAN512, a new survey for pulsars and transients with the Nan\c cay Radio Telescope, with observations made between 2012 and 2016. We described the processing pipeline, which is designed to remain sensitive to high-DM millisecond pulsars and to binaries with orbital periods $\gtrsim $ 3 h. The survey discovered one non-recycled pulsar and two MSPs, and redetected all previously known pulsars in the survey sky region bright enough for the sensitivity of the  survey. {Some unknown pulsars could have potentially been missed due to the incomplete survey.} We characterise the newly discovered MSP using Nan\c cay and Effelsberg radio data, and using gamma-ray data from the Large Area Telescope on the \textit{Fermi} satellite. 
A comparison of the yield of the  survey with preliminary population synthesis predictions suggests that far from the Galactic centre and the Galactic plane, the ratio of MSPs to non-recycled pulsars may be larger than currently believed. 

We timed PSR~J2205+6012 for 8 years with sub-microsecond precision. The \textsc{TempoNest} noise and timing analysis shows that the current timing precision is only limited by radiometer noise. {A TOA median uncertainty at 2.5\,GHz of 0.6\,$\mu$s and a timing rms of 0.75\,$\mu$s set this pulsar as one of the most precisely timed EPTA pulsars \citep{dcl+16,ccg+21}}. This promises
exciting prospects for future observations with ultra-wide-band receivers being commissioned at various telescopes in the Northern hemisphere such as Effelsberg and the Green Bank Telescope. Meanwhile, this dataset will be included in a future EPTA data release.
With the highly scattered pulse profile observed at L-band and the large DM variations inferred from our analysis, novel wide-band pulsar timing techniques such as \textsc{PulsePortraiture} \citep{pen19} or the Bayesian profile domain timing software \textsc{TempoNest2} \citep{lkd17+} will be of great interest.

\begin{acknowledgements}
The authors acknowledge the use of the MPIfR Hercules cluster hosted
at the Max Planck Computing and Data Facility in Garching and the
CNRS/IN2P3's Computing Centre. We also acknowledge support from the
CNRS/IN2P3 Computing Center (Lyon - France) for providing computing
and data-processing resources needed for this work. The Nan\c cay
Radio Observatory is operated by the Paris Observatory, associated
with the French Centre National de la Recherche Scientifique. This
publication is partly based on observations with the 100-m telescope of the MPIfR (Max-Planck-Institut f\"ur Radioastronomie) at Effelsberg.  This
research has made extensive use of NASA's Astrophysics Data System.

The \textit{Fermi} LAT Collaboration acknowledges generous ongoing support
from a number of agencies and institutes that have supported both the
development and the operation of the LAT as well as scientific data analysis.
These include the National Aeronautics and Space Administration and the
Department of Energy in the United States, the Commissariat \`a l'Energie Atomique
and the Centre National de la Recherche Scientifique / Institut National de Physique
Nucl\'eaire et de Physique des Particules in France, the Agenzia Spaziale Italiana
and the Istituto Nazionale di Fisica Nucleare in Italy, the Ministry of Education,
Culture, Sports, Science and Technology (MEXT), High Energy Accelerator Research
Organization (KEK) and Japan Aerospace Exploration Agency (JAXA) in Japan, and
the K.~A.~Wallenberg Foundation, the Swedish Research Council and the
Swedish National Space Board in Sweden.
Additional support for science analysis during the operations phase is gratefully
acknowledged from the Istituto Nazionale di Astrofisica in Italy and the Centre
National d'\'Etudes Spatiales in France. This work performed in part under DOE
Contract DE-AC02-76SF00515.

GD was supported by the European Research Council for the ERC Synergy
Grant BlackHoleCam under contract no. 610058. DAS thanks the International Space Science Institute (ISSI, Bern, Switzerland) for financial support of team 459 meetings that helped to improve the present work.
\end{acknowledgements}

\bibliographystyle{aa}
\bibliography{psrrefs,newrefs}


\end{document}